\documentclass[preprint2]{aastex}

\slugcomment{Accepted --- (20th September 2013)}

\shorttitle{Atmospheric production of cosmic rays}
\shortauthors{Obermeier et al.}

\begin{document}

\title{INTERACTIONS OF COSMIC RAYS IN THE ATMOSPHERE: GROWTH CURVES REVISITED}
\author{A. Obermeier\altaffilmark{1,2}, P. Boyle\altaffilmark{1,3}, J. H\"orandel\altaffilmark{2} and D. M\"uller\altaffilmark{1}}

\email{a.obermeier@astro.ru.nl}

\altaffiltext{1}{University of Chicago, Chicago, IL 60637, USA}
\altaffiltext{2}{Radboud Universiteit Nijmegen, 6525 HP Nijmegen, The Netherlands}
\altaffiltext{3}{Now at McGill University, Montreal, Canada}

\begin{abstract}
Measurements of cosmic-ray abundances on balloons are affected by interactions in the residual atmosphere above the balloon. Corrections for such interactions are particularly important for observations of rare secondary particles such as boron, antiprotons and positrons. These corrections can either be calculated if the relevant cross sections in the atmosphere are known, or may be empirically determined by extrapolation of the ``growth curves'', i. e.\ the individual particle intensities as functions of atmospheric depth. The growth-curve technique is particularly attractive for long-duration balloon flights where the periodic daily altitude variations permit rather precise determinations of the corresponding particle intensity variations. 
We determine growth curves for nuclei from boron ($Z=5$) to iron ($Z=26$), using data from the 2006 Arctic balloon flight of the TRACER detector for cosmic-ray nuclei, and we compare the growth curves with predictions from published cross section values. In general, good agreement is observed. We then study the boron/carbon abundance ratio and derive a simple and energy-independent correction term for this ratio. We emphasize that the growth-curve technique can be developed further to provide highly accurate tests of published interaction cross section values.
\end{abstract}
\keywords{balloon --- cosmic rays --- cross sections}

\section{Introduction}

This study is motivated by observations of high-energy cosmic-ray nuclei in long-duration balloon flights with the TRACER instrument~\citep{OberMeas}. Like all balloon-borne observations, the TRACER measurement was performed below a residual atmosphere of a few grams per cm$^2$. Thus, in order to determine the characteristics of the ambient cosmic-ray population on top of the atmosphere, the measurement must be corrected for changes due to particle interactions in the atmosphere. Subsequently, changes due to interactions in interstellar space must be taken into account in order to derive the properties of the cosmic-ray source.

Cosmic rays propagating through interstellar space may either diffuse to the edge of the Galactic disk (and halo), and then escape from the Galaxy, or they may be removed by spallation when colliding with a nucleus of the interstellar gas.  The diffusive escape is commonly characterized by introducing the propagation path length $L$ which is about 10~g~cm$^{-2}$ at energies of a few GeV~amu$^{-1}$, and appears to be independent of the chemical identity of the primary nucleus. It is well known that the path length $L(E)$ decreases with energy for highly relativistic nuclei (see e.g.~\citet{GP3} and references therein). Typical float altitudes of balloons correspond to a residual atmosphere of a few grams per cm$^2$. Thus, the amount of matter encountered by a cosmic-ray particle in the atmosphere is of the same order as that traversed in the ISM. However, the interstellar medium consists mostly of hydrogen (plus 10\% helium in number density), while oxygen and nitrogen are the main components of air. Due to this difference in composition, the interaction probability per unit mass density is larger in the ISM than in the atmosphere.

For a quantitative study of these effects, the appropriate interaction cross sections need to be known.  There are two kinds of cross sections: first, the total charge changing cross section describing the break-up of  a particular nucleus when interacting with a given target, and second, the ``differential'' cross section for a nucleus A to produce a specific nucleus B by spallation in a given target. It has been found long ago~\citep{BPform} that the total cross sections obey simple geometric scaling at high energies, and the parameterization given by~\citet{BPform} is still widely used~\citep{westfall,tsao93} and in good agreement with current data~\citep{webber90a,webber98a} and parameterizations~\citep{webber03}. The differential cross sections are more difficult to parameterize. A compilation of~\citet{webber03}, based on an interpolation of  accelerator measurements~\citep{webber90b,webber98a,webber98b}, is usually used for analysis of cosmic-ray data. As discussed by~\citet{AveInt}, the uncertainty in the values of the differential cross sections is larger than that for the total cross sections. There is insufficient experimental confirmation of the published cross section values, especially in the high-energy region of current cosmic-ray measurements. One usually assumes that the cross sections become independent of energy above a few GeV~amu$^{-1}$.

In the present study, we will use the TRACER 2006 balloon data~\citep{OberMeas} to investigate how empirically obtained ``growth curves'' in the atmosphere may provide constraints on the cross section values used in the analysis of the data. Growth curves describe the change in intensity of cosmic rays with atmospheric depth. 

This study is of specific concern for measurements of secondary cosmic-ray nuclei produced by spallation in the ISM. The abundances of these nuclei on top of the atmosphere, relative to those of their primary parents, are commonly taken to be a sensitive measure of the interstellar path length encountered by the primaries, and of the energy dependence of this path length. As a fairly well measured example, one often considers the relative abundances of boron and carbon (the ``B/C ratio'').  For balloon-borne measurements at sufficiently high energy, one must expect that the number of boron nuclei produced in the atmosphere equals or exceeds that of interstellar origin.  An accurate determination of the interstellar B/C ratio then critically depends on the knowledge of the atmospheric correction for boron, i.e.\ of the production cross sections for boron in the atmosphere (for details, see~\citet{OberInt}). 
\begin{figure}[t]
 \includegraphics[width=\linewidth]{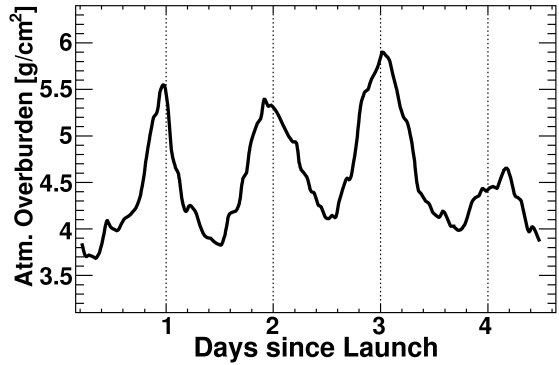}
 \caption{Vertical atmospheric overburden as a function of flight time for the TRACER 2006 balloon flight.} \label{fig1}
\end{figure}

\section{INTERACTIONS OF PRIMARY COSMIC RAYS IN THE ATMOSPHERE}

Long-duration balloon flights are now routinely performed with zero-pressure balloons at polar latitudes and during local summer. The extended duration of such flights, up to several weeks, is made possible by the fact that the sun remains above the horizon for the entire flight. Nevertheless, the zenith angle of the Sun undergoes diurnal variations, leading to periodic changes in the float altitude of the balloon. These variations permit measurements of the particle intensities as a function of residual overburden with higher accuracy than possible with a traditional measurement, performed during just one short balloon ascent. This is illustrated in Figure~\ref{fig1} for 4.5~days of flight of TRACER in 2006. The minimum vertical atmospheric overburden $X$ for this flight is 3.7~g~cm$^{-2}$, and the maximum 5.8~g~cm$^{-2}$. Within these limits, the detector continuously scans the intensities $N(X)$ for all components. The function $N(X)$ is called the growth curve. The growth curve can be used to extrapolate the measurement to the intensity $N(0)$ on top of the atmosphere. 

\begin{figure}[t]
 \includegraphics[width=\linewidth]{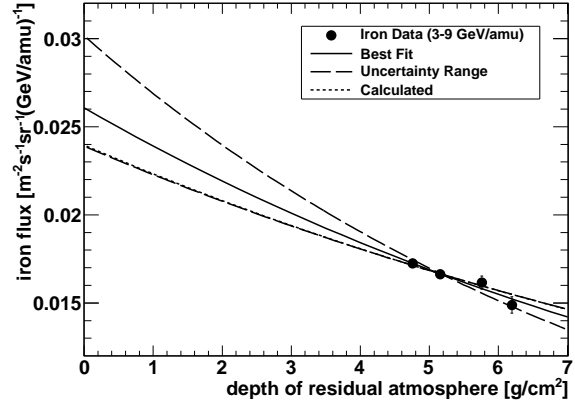}
 \caption{Growth curve of iron. The best model fit is shown (solid) with uncertainty ranges (dashed) and compared to the curve using the calculated interaction path length (dotted). Note that the dotted line (calculation) lies atop the lower dashed line (fit uncertainty).}\label{fig2}
\end{figure}

Measured growth curves from the 2006 TRACER flight for iron, oxygen and carbon nuclei in the energy interval from 3 to 9~GeV~amu$^{-1}$ are shown in Figures~\ref{fig2}, \ref{fig3} and~\ref{fig4}. It should be noted that the measurement for each event determines the angle of incidence of the cosmic-ray particle. Hence, the abscissa in these figures represents the actual atmospheric depth traversed by the particle, which in general is slightly larger than the vertical overburden above the detector. In first order, one may just linearly extrapolate the curves to $X=0$ and obtain the intensities of these elements at the top of the atmosphere, provided that the absolute intensities are properly normalized. The uncertainties in the extrapolation are of the order of a few percent. At higher energies, the decreasing counting rate will reduce the accuracy of this procedure. The atmospheric correction of the measurements must then be based on calculations using independent information on the relevant cross sections.

For an abundant primary component, such as iron, the growth curve $N_{Fe}(X)$ is dominated by losses due to spallation of iron nuclei in the atmosphere, characterized by an attenuation path length $\Lambda_{Fe}$:
\begin{eqnarray}
\label{Eq:Iron}
N_{Fe}(X)  &=& N_{Fe}(0)\cdot  \exp(-X/\Lambda_{Fe}) \nonumber \\
&\approx & N_{Fe}(0) (1-X/\Lambda_{Fe}),
\end{eqnarray}
with $\Lambda_{Fe} = m/\sigma_{Fe}$ , and $m$ and $\sigma$ the average nuclear target mass and spallation cross section, respectively.

An analogous expression is valid for oxygen nuclei. While there may also be production of secondary oxygen by spallation of heavier elements, we may ignore this contribution as these elements are much less abundant than oxygen.

For carbon a small contribution of secondary carbon, produced in the atmosphere by spallation of oxygen, must be included:
\begin{eqnarray}
\label{Eq:Carbon}
N_C(X) &=& N_C(0)\cdot \exp(-X/\Lambda_C) \nonumber \\
&+& N_O(0)[1-\exp(-X/\lambda_{O\rightarrow C})].
\end{eqnarray}
Here, the spallation of oxygen into carbon is parameterized by the ''differential path length'' $\lambda_{O\rightarrow C}$.
A first-order linearization of Eq.~(\ref{Eq:Carbon}), and the assumption of nearly equal abundances of oxygen and carbon, $N_C(0) \approx N_O(0)$~\citep{OberMeas}, lead to
\begin{eqnarray}
\label{Eq:ApproxCarbon}
N_C(X) &\approx & N_C(0) (1-X/\Lambda_C) + N_O(0) X/\lambda_{O\rightarrow C} \nonumber \\
 & \approx & N_C(0) (1-X(\Lambda_C^{-1}  -  \lambda_{O\rightarrow C}^{-1})).
\end{eqnarray}
\begin{equation}
\label{Eq:ApproxCarbon2}
N_C(X) \approx N_C(0) (1-X/\Lambda_C^{\prime})
\end{equation}
The attenuation path lengths $\Lambda_{Fe}$, $\Lambda_O$ and $\Lambda_C$ can be calculated from the the total charge-changing cross sections mentioned above~\citep{webber90a}. As these cross sections are determined in carbon targets, they need to be scaled to air (79\%~N$_2$ and 21\%~O$_2$ by number) using the proportionality to $A^{2/3}$~\citep{BPform,westfall}. A conservative uncertainty of 25\% is estimated for the cross sections due to the scaling, assumption of energy independence (measurements are only available below 3~GeV~amu$^{-1}$) and differences of the values given by~\citet{webber90a} to other measurements and calculations (e.g. \citet{read84, korejwo00, tsao98}). The interaction path lengths derived are then: 
\begin{figure}[t]
 \includegraphics[width=\linewidth]{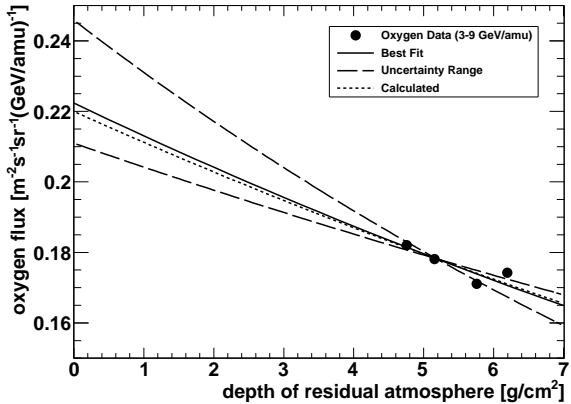}
 \caption{Growth curve of oxygen. The best model fit is shown (solid) with uncertainty ranges (dashed) and compared to the curve using the calculated interaction path length (dotted).}\label{fig3}
\end{figure}
\begin{eqnarray}
\label{Eq:Lit}
&\Lambda_{Fe}&  = 14.2\pm1.8 \, \rm{g/cm}^2, \nonumber \\
&\Lambda_O & =  24.6\pm3.1\, \rm{g/cm}^2, \nonumber \\
&\rm{ and }&\Lambda_C  = 27.4\pm3.4\, \rm{g/cm}^2.
\end{eqnarray}
The value of  $\lambda_{O\rightarrow C}$ requires knowledge of the differential cross section $\sigma_{O\rightarrow C}$. Using a value of $\sigma_{O\rightarrow C} = 123.2$~mb ($\pm1.5$\%) from the compilation of~\citet{webber90b}, we obtain, after scaling from carbon target to air, an apparent path length $\Lambda_C^\prime$ for carbon (as defined in Equation~\ref{Eq:ApproxCarbon2}) that is slightly larger than $\Lambda_C$:
\begin{equation}
\label{Eq:Litapprox}
\Lambda_C^\prime \approx \Lambda_C\cdot \frac{\lambda_{O\rightarrow C}}{\lambda_{O\rightarrow C} - \Lambda_C} = 30.7\pm3.8\,\rm{g/cm}^2.
\end{equation}
One may also determine the attenuation path lengths directly from the measured growth curves (Figures~\ref{fig2}, \ref{fig3}, and~\ref{fig4}). The figures include fits to data corresponding to Equations (\ref{Eq:Iron}) and (\ref{Eq:ApproxCarbon2}) with $\Lambda_{Fe}$, $\Lambda_O$,  and  $\Lambda_C^\prime$ as free parameters, respectively. The uncertainty ranges of the fits are included as dashed lines. The empirical values according to these fits are $\Lambda_{Fe}=11.5\pm2.8$~g~cm$^{-2}$, $\Lambda_O=23.4\pm7.3$~g~cm$^{-2}$, and $\Lambda_C^\prime=38.6\pm8.4$~g~cm$^{-2}$. The empirical values are in good agreement with the calculated values of Equation (\ref{Eq:Lit}) and (\ref{Eq:Litapprox}) within the measurement uncertainties. A summary of the calculated and measured interaction path lengths is given in Table~\ref{tab:lambdas}.  This agreement confirms the validity of the cross section values obtained from the literature, but also, justifies the use of empirical growth curves to derive atmospheric corrections.

\begin{figure}[t]
 \includegraphics[width=\linewidth]{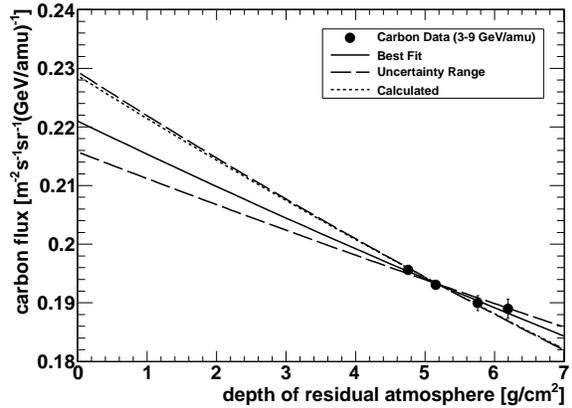}
 \caption{Growth curve of carbon. The best model fit is shown (solid) with uncertainty ranges (dashed) and compared to the curve using the calculated interaction path length (dotted). Note that the dotted line (calculation) lies atop the upper dashed line (fit uncertainty).}\label{fig4}
\end{figure}

In principle, it should be possible to use the empirical growth curves to determine values for the spallation cross sections with an accuracy that exceeds that of the parameterizations found in the literature.  However, the interpretation of the TRACER data is limited in counting statistics, and may be affected by drifts in detector efficiency. These limitations can be remedied in a flight of longer duration (more than 20 days, and hence, more than 20 cycles in the altitude profile, are now almost routinely achieved in Antarctica\footnote{The recent Antarctic balloon flight of the Super-TIGER experiment had a record duration of 55 days.}), and  with a dedicated effort to monitor drifts in detector response. An accuracy around 1\% in determining cross sections from the measured growth curves should then be within reach.

\section{GROWTH CURVE FOR BORON, AND THE B/C RATIO}

The intensity profile of boron in the atmosphere is affected by the loss of Galactic boron due to interactions, and by the production of secondary boron from spallation of heavier cosmic-ray nuclei in the atmosphere. The dominant parent nuclei for such spallations, carbon and oxygen, are much more abundant than boron. Hence, the spallation production offsets much of the spallation losses in the atmosphere. The growth curve has the form
\begin{eqnarray}
\label{Eq:Boron}
N_B(X) &=& N_B(0)\cdot \exp(-X/\Lambda_B) \nonumber \\
&+& \sum N_k(0)[1-\exp(-X/\lambda_{k\rightarrow B})].
\end{eqnarray}
The summation includes all cosmic rays heavier than boron, with $\lambda_{k\rightarrow B}$ characterizing the differential  path length  for spallation into boron. In practice, only carbon and oxygen contribute significantly, and we may  linearize the spallation production:
\begin{equation}
\label{Eq:BoronApprox}
N_B(X)  \approx N_B(0)\cdot \exp(-X/\Lambda_B) + N_C(0)\cdot\frac{X}{\lambda_{eff}}
\end{equation}
with
\begin{equation}
\label{Eq:LamEff}
\frac{1}{\lambda_{eff}} = \frac{1}{\lambda_{C\rightarrow B}} + \frac{N_O(0)}{N_C(0)}\cdot\frac{1}{\lambda_{O\rightarrow B}}.
\end{equation}
The effective differential spallation path length $\lambda_{eff}$ is determined by the spallation path length of carbon to boron, and by the spallation of oxygen. Once again, one may accept $N_C(0)/N_O(0) \approx 1$.

\begin{table}[t]
\begin{center}
\caption{Summary of the measured and calculated interaction path lengths in~g~cm$^{-2}$. See text for details.}
\label{tab:lambdas}
\begin{tabular}{cccc}
\tableline\tableline
Element & Quantity  & measured & calculated\\
\tableline
Fe & $\Lambda_{Fe}$, Eq.(\ref{Eq:Iron}) & $11.5\pm2.8$ & $14.2\pm1.8$ \\
O & $\Lambda_{O}$, Eq.(\ref{Eq:Iron}) & $23.4\pm7.3$ & $24.6\pm3.1$ \\
C & $\Lambda_{C}^\prime$, Eq.(\ref{Eq:ApproxCarbon2}) & $38.6\pm8.4$ & $30.7\pm3.8$ \\
B & $\lambda_{eff}$, Eq.(\ref{Eq:LamEff}) & $225\pm102$ & $227\pm 44$ \\
\tableline
\end{tabular}
\end{center}
\end{table}

Using tabulated cross sections from the literature, we calculate $\Lambda_B = 28.7\pm3.7$~g~cm$^{-2}$, and $\lambda_{eff} = 227\pm 44$~g~cm$^{-2}$. The uncertainty of $\lambda_{eff}$ originates from the uncertainty in the assumption $N_C(0)=N_O(0)$, and from a conservative estimate of the errors in $\lambda_{C\rightarrow B}$ and $\lambda_{O\rightarrow B}$ due to the applied scaling and differences in the published results of~\citep{webber90a,read84,korejwo00,tsao98}.
 
\begin{figure}[t]
 \includegraphics[width=\linewidth]{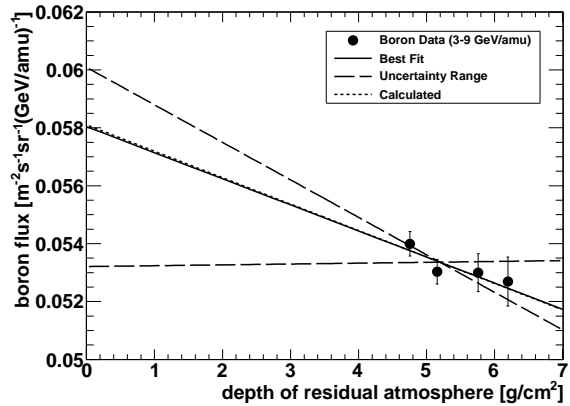}
 \caption{Growth curve of boron. The best model fit is shown (solid) with uncertainty ranges (dashed) and compared to the curve using the calculated interaction path length (dotted).  Note that the dotted line (calculation) lies atop the solid line (best fit).}\label{fig5}
\end{figure}

In Figure~\ref{fig5}, we show the measured growth curve for boron. In order to obtain the path length parameters corresponding to Equation~(\ref{Eq:BoronApprox}), the intensity $N_C(0)$ must be known.  We accept the results from the HEAO-3 measurement~\citep{heao} which have provided highly accurate results in the 3 to 9~GeV amu$^{-1}$ region. The results from TRACER~\citep{OberMeas} also cover this energy region and agree well with HEAO-3. With a calculated path length $\Lambda_B = 28.7$~g~cm$^{-2}$, the fitted curves shown in Figure~\ref{fig5} correspond to a fitted value $\lambda_{eff} = 225\pm102$~g~cm$^{-2}$. This value is in remarkable agreement with the calculated number given above, in spite of the large uncertainties (Table~\ref{tab:lambdas} summarizes the results).
Once again, we conclude that the absorption and secondary production of cosmic rays deduced from atmospheric growth curves is in good agreement with expectations using cross section values available in the literature.

\section{THE B/C RATIO AT HIGH ENERGIES}

One of the major goals of measurements such as that performed with TRACER is to determine the energy dependence of the galactic propagation path length $L(E)$ at high energies, around 1000~GeV~amu$^{-1}$ or higher. Will $L$ continue to decrease even at the highest energies, or will the path length reach a limiting final value that is independent of energy? In any case, correction of the data for atmospheric secondaries will become very important, but the data will likely be severely limited by decreasing statistics at high energy. Thus, we need to predict the magnitude of the correction from the observed behavior at low energy.
The intensity of boron on top of the atmosphere, $N_B(0)$,  from Equations~(\ref{Eq:Boron}) and~(\ref{Eq:BoronApprox}) is
\begin{eqnarray}
\label{Eq:Babundance}
N_B(0) &\approx & N_B(X)\cdot \exp(X/\Lambda_B) \nonumber \\
&-& N_C(0)\frac{X}{\lambda_{eff}}\cdot \exp(X/\Lambda_B),
\end{eqnarray}
and the B/C ratio on top of the atmosphere becomes
\begin{eqnarray}
\label{Eq:B/C}
\frac{N_B(0)}{N_C(0)}&\approx &  \frac{N_B(X)\cdot \exp (X/\Lambda_B)}{N_C(0)} \nonumber \\
 &-& X/\lambda_{eff}\cdot \exp(X/\Lambda_B). 
\end{eqnarray}
The first term in Equation~(\ref{Eq:B/C}) is the apparent B/C ratio on top of the atmosphere if one were to ignore the boron production in the atmosphere but included the small atmospheric correction for carbon as described in Equations~(\ref{Eq:Carbon}) and~(\ref{Eq:ApproxCarbon}). The atmospheric production of boron is taken into account with the second term in Equation~(\ref{Eq:B/C}). This term is proportional to the atmospheric depth $X$, but does not depend on energy. Thus, for a balloon borne measurement, this correction term could be applied at all energies.

One may note that this conclusion depends on two assumptions: (a) the spallation cross sections are assumed to be independent of energy, and (b) the intensity ratio of the parent nuclei carbon and oxygen is taken as energy-independent and roughly equal to unity.  The C/O ratio, however, decreases slightly with energy as the spallation fraction of Galactic carbon decreases. A change in the C/O ratio by 5\% would alter the result for the production path length of boron in the atmosphere by about 1.5\%.

\begin{figure}[t]
 \includegraphics[width=\linewidth]{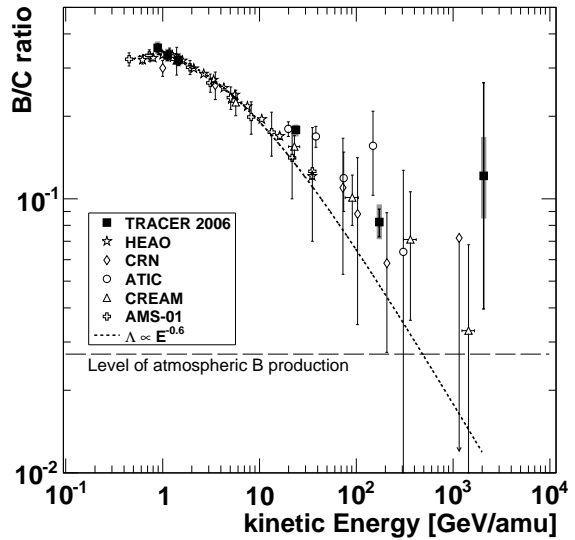}
 \caption{Boron to Carbon ratio~\citep{OberMeas}. The dotted line represents a simple model of the cosmic-ray escape path length proportional to $E^{-0.6}$. The dashed line indicated the contribution of boron produced in the atmosphere (Eq.~\ref{Eq:BCcorrection}).  } \label{fig6}
\end{figure}

To illustrate the magnitude of the atmospheric correction, we show in Figure~\ref{fig6} a compilation of B/C data from several experiments. Except for HEAO-3, CRN, and AMS-01, these measurements have been made on balloons, and atmospheric corrections reportedly have been made. For the TRACER flight, at an average altitude of  $X = 5.2$~g~cm$^{-2}$, the correction term amounts to
\begin{equation}
\label{Eq:BCcorrection}
X/\lambda_{eff}\cdot \exp(X/\Lambda_B) =  0.027^{+0.007}_{-0.004}.
\end{equation}
The level of the subtracted atmospheric correction for all energies and for the same atmospheric overburden as that of TRACER  is indicated by the horizontal line in Figure~\ref{fig6}. The statistical uncertainties in the B/C measurements at high energy are quite large. If the ratio continued to fall steadily with energy, the corrected ratio would equal the magnitude of the correction at energies around 1000~GeV~amu$^{-1}$. However, it cannot be excluded that the ratio flattens out to a constant value at those energies that well exceeds the magnitude of the correction.  The astrophysical implications of such a behavior would, of course, be highly interesting.

\section{Conclusion}

In this study, we have utilized a specific feature in the conduct of cosmic-ray measurements with zero-pressure balloons in long-duration flights at Arctic or Antarctic latitudes, namely the fact that the balloon altitude undergoes diurnal variations corresponding to variations of several g~cm$^{-2}$ in the residual atmosphere. This feature makes it possible to determine the growth curves of cosmic-ray intensities with higher accuracy than typically obtained in conventional balloon flights and allows to determine cosmic-ray fluxes at the top of the atmosphere solely with cosmic-ray data collected during flight. We have studied a few consequences derived from this feature.

We have determined the intensities of the secondary cosmic-ray nucleus boron, and of the primary nuclei carbon, oxygen, and iron on top of the atmosphere by extrapolation of the empirical growth curves measured in the 2006 flight of the TRACER instrument. We have found that the accuracy of the extrapolated intensities is of the order of a few percent when the measured data are taken at relatively low energies (3-9~GeV~amu$^{-1}$), and hence, have good counting statistics.

The extrapolation from flight altitude to the top of the atmosphere can also be performed with the use of numerical cross section values. This procedure generally leads to good agreement with the growth-curve technique.

This study was motivated by the great interest in measurements on balloons to obtain accurate data on the B/C ratio at high energies. Inevitably, these measurements become more and more statistics-limited as the energy increases. Thus, it is very important that the atmospheric correction for the B/C ratio can be derived from a simple, energy-independent expression. Such an expression is given in Equation~(\ref{Eq:B/C}). The contribution of the atmosphere to the measured B/C ratio is about 0.5\%~per~g~cm$^{-2}$ of residual atmosphere (Equation~(\ref{Eq:BCcorrection})). Accurate knowledge of this contribution makes it possible to obtain reliable information on the B/C ratio even if the atmospheric contribution to boron exceeds the magnitude of the Galactic intensity.

Finally, it should be emphasized that the growth-curve technique in long-duration balloon flights may also have great utility for other measurements of rare cosmic-ray components, such as rare isotopes (e.g. $^{10}$Be), heavy nuclei, and positrons and anti-protons. 

\acknowledgments
AO acknowledges support of FOM in the Netherlands (``Stichting voor Fundamenteel Onderzoek der Materie''). This work was supported by NASA through grants NNG 04WC08G, NNG 06WC05G, and NNX 08AC41G.

\end{document}